\documentclass[a4paper,12pt]{article}

\usepackage{amsmath}\usepackage{amssymb}
\usepackage{latexsym}\usepackage{cite}

\usepackage[dvips]{graphicx}
\usepackage{pstricks}
\usepackage{bm}

\usepackage{graphicx}

\usepackage{amsmath, amssymb, graphics}

\newcommand{\mathsym}[1]{{}}

\newcommand{\be}{\begin{equation}}
\newcommand{\ee}{\end{equation}}

\newcommand{\ka}{\kappa}

\def\beq{\begin{equation}}
\def\eeq{\end{equation}}
\def\beqr{\begin{eqnarray}}
\def\eeqr{\end{eqnarray}}

\def\al{\alpha}
\def\bt{\beta}
\def\Ga{\Gamma}
\def\ga{\gamma}
\def\de{\delta}
\def\De{\Delta}

\def\ka{\kappa}
\def\si{\sigma}

\def\te{\theta}

\def\La{\Lambda}
\def\lam{\lambda}

\def\om{\omega}
\def\ep{\epsilon}
\def\vep{\varepsilon}
\def\vp{\varphi}

\def\l{\left (}
\def\r{\right )}

\def\fr{\frac}
\def\la{\label}
\def\hs{\hspace}
\def\vs{\vspace}

\def\ran{\rangle}
\def\lan{\langle}
\def\ov{\overline}
\def\tl{\tilde}
\def\tm{\times}

\begin{document}

\begin{flushright}
September 28, 2022 \\
\end{flushright}

\vs{1.5cm}

\begin{center}
{\Large\bf

SM extension with a gauged flavor $U(1)_F$ symmetry}

\end{center}

\vspace{0.5cm}
\begin{center}
{\large
{}~Zurab Tavartkiladze\footnote{E-mail: zurab.tavartkiladze@gmail.com}
}
\vspace{0.5cm}

{\em Center for Elementary Particle Physics, ITP, Ilia State University, 0162 Tbilisi, Georgia}

\end{center}
\vspace{0.6cm}

\begin{abstract}


An extension of the Standard Model with anomaly free $U(1)_F$ flavor symmetry is studied in this paper.
With this extension and  the addition of the right-handed neutrino states, the solution of
anomaly free charge assignments is found, which gives
 appealing texture zero and hierarchical Yukawa matrices. This gives us a natural understanding
 of the hierarchies between charged fermion masses and Cabibbo-Kobayashi-Maskawa (CKM) matrix elements.
  Neutrino Dirac and Majorana coupling matrices also have desirable structures
 leading to successful neutrino oscillations with inverted neutrino mass ordering. Other
 interesting implications of the presented scenario are also discussed.

\end{abstract}

\hspace{0.4cm}{\it Keywords:}~Flavor symmetry; Fermion masses; Neutrino oscillations.

\hspace{0.4cm}PACS numbers:~11.30.Hv, 12.15.Ff, 14.60.Pq





\section{INTRODUCTION}

Although being very successful, the Standard Model (SM) is unable to resolve some puzzles.
Among them is a problem of fermion flavor. The origin of hierarchies between charged fermion masses and CKM mixing
angles is unexplained. Moreover, the SM is unable to accommodate
the neutrino data \cite{Capozzi:2021fjo}.
In this work we consider an extension that gives a simultaneous resolution of these problems.
The extension, we consider is the flavor $U(1)_F$ symmetry, which will be gauged. Besides this, we augment the fermion
sector with right-handed neutrinos (RHNs), which will be responsible for the generation of light neutrino masses and mixings.

While the Abelian flavor $U(1)_F$ is the simplest candidate for the flavor symmetry  \cite{Froggatt:1978nt}, its gauging  is a challenging
task because the anomaly cancelation conditions give severe constraints for realistic model building.
Below we present our findings of the $U(1)_F$ charge assignment.

\section{ANOMALY FREE FLAVOR $U(1)_F$}

Earlier attempts to find an anomaly free setup with $U(1)_F$ symmetry,
exist in the literature \cite{Dudas:1995yu, Chen:2008tc,Tavartkiladze:2011ex}.
These have been  either within the minimal supersymmetric extension of the SM
 \cite{Dudas:1995yu} or within the supersymmetric grand unified theories (GUTs) \cite{Chen:2008tc,Tavartkiladze:2011ex}.
In \cite{Tavartkiladze:2011ex}  for the finding of the anomaly free  $U(1)_F$ symmetries, extended GUT  symmetry groups
[unifying $SU(5)$ GUT and $U(1)_F$ (or some part of the latter)] has been used. Although this approach is very attractive, with
unification putting additional constraints, it disallows us to have much texture zeros and predictions. Besides these, GUTs usually suffer
other problems that are not directly related to the flavor symmetry.
Since we feel that finding anomaly free $U(1)_F$ constructions is far from being fully explored, our study here
will be SM extension with gauged $U(1)_F$ symmetry and RHN states.

The nontrivial  states under the SM gauge group $G_{SM}=SU(3)_c\tm SU(2)_L\tm U(1)_Y$ that we introduce will be just those of
the SM. These are the Higgs doublet $\vp $
and three families of matter $\{q, u^c, d^c, l,e^c \}_{i=1,2,3}$
 where $i=1,2,3$ is the family index.\footnote{
Here and below for fermionic $f$ states we use  a two-component Weyl spinor in $(\fr{1}{2},0)$ representation of the
Lorentz group.}
As far as the extension is concerned, the fermionic sector will be augmented with RHNs  $N_{1,2,3\cdots }$.
As already emphasized, the extra gauge symmetry $U(1)_F$ is considered, with the scalar field $X$ - the "flavon" - needed for the $U(1)_F$
breaking.

  For finding anomaly free $U(1)_F$
charges we will use several simple observations. First of all, recall that the simplest anomaly free $U(1)$ symmetry is the hypercharge symmetry $U(1)_Y$ 
- the part of the SM gauge sector. So, in principle for  $U(1)_F$, the family dependent hypercharges can be used. Furthermore,
by introducing the right-handed neutrinos  one can also build the gauged $(B-L)$ symmetry, which is also anomaly free.
 Obviously, with family dependent $(B-L)$ charges, anomalies will still vanish.
So, one option is to have $U(1)_F$'s charges $Q_i(f)$ as the following superposition
$\bar a_iY(f)+\bar b_iQ_{B-L}(f)$, where $\bar a_i, \bar b_i$ are some constants. With this superposition,  all anomalies of the $G_{SM}$ remain intact, and
also following additional and mixed anomalies
\begin{subequations}
\begin{align}
(U(1)_F)^3:&~~~~A_{111}=\sum_i Q_i^3 ~,\label{A111}\\
U(1)_Y\tm (U(1)_F)^2:&~~~~A_{Y11}=\sum_i Y_iQ_i^2~, \label{AY11} \\
(U(1)_Y)^2\tm U(1)_F:&~~~~A_{YY1}=\sum_i Y_i^2Q_i ~,\label{AYY1} \\
(SU(2)_L)^2\tm U(1)_F:&~~~~A_{221}=\sum_i \left [ Q_i(l_i)+3Q_i(q_i)\right ]~, \label{A221}  \\
(SU(3)_c)^2\tm U(1)_F:&~~~~A_{331}=\sum_i \left [2Q_i(q_i)+Q_i(u^c_i)+Q_i(d^c_i)\right ] ~, \label{A331} \\
({\rm Gravity})^2\tm U(1)_F:&~~~~A_{GG1}=\sum_i Q_i~ .\label{AGG1}
 \end{align}
\end{subequations}
automatically vanish.
Although, for the $Q_i(f)$ assignments, the superposition   $\bar a_iY(f)+\bar b_iQ_{B-L}(f)$ can be considered,
one immediate outcome is that, by requiring that the top quark has
the renormalizable Yukawa coupling ($\lam_t\sim 1$) with the Higgs doublet $\vp $, the bottom quark and the tau lepton Yukawas
will be allowed at renormalizable levels - with the expectancy that $\lam_b, \lam_{\tau}\sim 1$. Besides this unpleasant fact, with only $\bar a_i, \bar b_i$
we cannot get Yukawa coupling matrices with many texture zeros. Thus, for the $U(1)_F$'s charges we will consider the modified superposition
\beq
Q_i(f)=\bar a_iY(f)+\bar b_iQ_{B-L}(f)+\De Q_i(f)~,
\la{QU1F}
\eeq
where the additions $\De Q_i(f)$ will be selected in such a way that the anomalies $A_{YY1}, A_{221}, A_{331}, A_{GG1}$ stay intact.
However, for the vanishing of the anomalies $A_{111}$ and $A_{Y11}$ additional constraints on the charge prescriptions need to be imposed.
It turns out that for this goal, instead of three RHNs, we will need four of them - $N_{1,2,3,4}$.
The additions that satisfy these and give a desirable fermion pattern are
\begin{subequations}
\begin{align}
\De Q_i(q)&=\bar q_3\{0, 1, -1 \}+\bar q_8\{1, 1, -2 \}~,  \label{Deq}\\
\De Q_i(u^c)&=\bar u_3\{0, 1, -1 \}+\bar u_8\{1, 1, -2 \}~, \label{Deu} \\
\De Q_i(d^c)&=\bar d_3\{1, -1, 0 \}+\bar d_8\{1, 1, -2 \}~, \label{Ded} \\
\De Q_i(l)&=\bar l_3\{1, -1, 0 \}+\bar l_8\{1, 1, -2 \}~, \label{Del}  \\
\De Q_i(e^c)&=0~, \label{Dee} \\
\De Q_i(N)&=\bar n\{1, 1, 1, -3 \}~ ,\label{DeN}
 \end{align}
 \la{Deltas}
\end{subequations}
\!where $\{\cdots \}$ stand for diagonal matrices in flavor space and presented numbers are diagonal elements of the corresponding matrices.
Note that, being traceless, these additions coincide with diagonal (Cartan) generators of $SU(3)$ [in Eqs. (\ref{Deq})-(\ref{Del})] and $SU(4)$
unitary groups [in Eq. (\ref{DeN})]. Thus the notations for
the constants $(\bar q_{3,8},\cdots \!,\bar l_{3,8})$  become obvious. These constants, together with $\bar a_i, \bar b_i$ will be enough
for our purposes.
Upon selecting these constants we will bear in mind some
 requirements that need  to be satisfied in order to obtain desirable and phenomenologically viable model.
These requirements are as follows:

 {\bf (i)} In order to have top quark Yukawa coupling $\lam_t\sim 1$, the $U(1)_F$ symmetry should allow coupling $q_3u^c_3\vp $ at a renormalizable level.
At the same time, all other Yukawa terms (responsible for charged fermion masses)
should emerge by spontaneous breaking of the $U(1)_F$.
So, the adequate mass hierarchies and CKM mixings will be expressed by powers of $\lan X\ran/M_{\rm Pl}$.

 {\bf (ii)} Dirac and Majorana-type couplings involving RHN $N$ states should be such that naturally generate light neutrino masses and mixings
 in order to accommodate recent neutrino data \cite{Capozzi:2021fjo}.

{\bf (iii)} While the  $U(1)_F$ charge assignment ansatz of Eqs. (\ref{QU1F}),  (\ref{Deltas}) automatically ensure zero anomalies of
(\ref{AYY1})-(\ref{AGG1}), an additional constraints need to be imposed for canceling anomalies of (\ref{A111}) and (\ref{AY11}).

{\bf (iv)} Finally, the ratios of the states' charges should be rational in order to allow (phenomenologically required) couplings between them.

Guided by these, in (\ref{QU1F}) we use normalization such that $Y(l)=1$ and $Q_{B-L}(q)=1/3, Q_{B-L}(l)=-1$.
Also, without loss of any generality, for the scalar $X$, we will select $Q_X=1$.
With these and requirements listed above, the best selection that we find is the following\footnote{
Other solutions, we found, either do not give desirable hierarchies for the whole fermion sector (including neutrinos), or
in some part do not work at all. We do not find it worthy to present such possibilities in this
 work; we give only one solution, which does not have any drawback.}:
$$
\bar a_i=\fr{1}{3}\{46, 43, 10 \}~,~~~\bar b_i=\fr{1}{3}\{-91, 35, 38 \}~,~~~
$$
$$
\{ \bar q_3, \bar u_3,\bar d_3,\bar l_3\}=\fr{1}{3}\{-16, 7, -67/2 ,-3/2\}~,
$$
\beq
\{ \bar q_8, \bar u_8,\bar d_8,\bar l_8\}=\fr{1}{9}\{38, -41, 23/2 ,51/2\}~,~~~~\bar n=-\fr{5}{3}~.
\la{ab38-select}
\eeq
With these, by using (\ref{QU1F}) and (\ref{Deq})-(\ref{DeN}) we obtained the charges given in Table \ref{t:tab1}. One can readily check out
that all anomalies given in (\ref{A111})-(\ref{AGG1}) vanish.
Note that after all charges are fixed, since whole Lagrangian respects $U(1)_Y$ symmetry, by making a family universal charge shift for the states
 $Q\to Q+\al Y$,
all couplings and anomalies will remain intact. The constant $\al $ can be selected to have convenient form of the charges. We have already
exploited this by setting $Q(q_3)=0$ (see Table \ref{t:tab1}).
 Presented charge assignment give interesting textures for charged fermion mass matrices and neutrinos
as well. These we discuss in the following sections.

\section{QUARK AND CHARGED LEPTON YUKAWA TEXTURES}

As mentioned, for the $U(1)_F$ gauge symmetry breaking, the SM singlet scalar $X$ - the flavon field - is introduced
and its $U(1)_F$ charge is taken to be $Q_X=1$.
The vacuum expectation value (VEV) $\lan X\ran $ breaks the $U(1)_F$ and also forms fermion mass matrices.
Since in the Yukawa couplings the appropriate powers of $\fr{X}{M_{\rm Pl}}$ and $\fr{X^*}{M_{\rm Pl}}$ will
appear, it is convenient to introduce notations
\beq
\fr{X}{M_{\rm Pl}}\equiv \vep ~,~~~~\fr{X^* }{M_{\rm Pl}}\equiv \bar{\vep }~.
\la{eps-epsbar1}
\eeq
 Note that $M_{\rm Pl}\simeq 2.4\tm 10^{18}$~GeV is a reduced Planck scale, which will be treated as a natural cutoff for all higher dimensional 
 nonrenormalizable operators.

With the $U(1)_F$ charges of the Higgs doublet $\varphi$ of  $Q_{\varphi }=-7$, and of the fermion states
given in Table \ref{t:tab1},
the $qu^c\varphi, qd^c\tl \varphi $ and $le^c\tl \varphi $ type couplings,  involving different powers of
$\vep$ and $\bar{\vep }$, will be:
%
%
%
%
\begin{table}
\caption{$U(1)_F$ charge ($Q$) assignment for the states. $Q_X=1$, $Q_{\vp }=-7$.
 }

\label{t:tab1} $$\begin{array}{|c||c|c|c|c|c|c|}

\hline
\vs{-0.3cm}
 &  &  &  &  &  &   \\

\vs{-0.4cm}

& \{q_1, q_2, q_3\}& \! \{u^c_1, u^c_2, u^c_3\}\!&
\! \{d^c_1, d^c_2, d^c_3\}\!&\! \{l_1, l_2, l_3\}\! & \! \{e^c_1, e^c_2, e^c_3\}\! & \!\!\{N_1, N_2, N_3, N_4\}\!\!\\

&  &  &  &  &  & \\

\hline

\vs{-0.3cm}
 &  &  &  &  &  & \\

\vs{-0.3cm}
\hs{-0.5mm}Q\hs{-0.5mm}&\!\hs{-0.7mm} \{-11, -2, 0\} \hs{-0.7mm}\!& \hs{-0.7mm} \{26, 13, 7\} \hs{-0.7mm}  &\!\hs{-0.7mm} \{\!-10,\!-1,\!-9\} \hs{-0.7mm}\!  &\!\{48, 6, -15 \}\! &\!\!\{-61,\!-17,6\}\!\!&\{ -32, 10, 11, 5\}\\

&  &  &  &  &  &\\

\hline

\end{array}$$

\end{table}
%
%
%
%
$$
\left(\!\!
  \begin{array}{ccc}
    q_1, &\!\!q_2, &\!\!q_3 \\
  \end{array}\!\!\right)
  \!\!\left(\!
    \begin{array}{ccc}
      \ov \vep^8 & \vep^5 & \vep^{11} \\
     \ov \vep^{17} &\ov \vep^4 & \vep^2 \\
      \ov \vep^{19} & \ov \vep^6 & 1 \\
    \end{array}\!\right)
    \!\!\left(\!\!
      \begin{array}{c}
        u^c_1 \\
        u^c_2 \\
        u^c_3 \\
      \end{array}\!\!\right)\!\!\varphi +
   \left(\!\!
  \begin{array}{ccc}
    q_1, &\!\!q_2, &\!\!q_3 \\
  \end{array}\!\!\right)
  \!\!\left(\!
    \begin{array}{ccc}
       \vep^{14} & \vep^5 & \vep^{13} \\
      \vep^5 & \ov \vep^4 &\vep^4 \\
      \vep^3 & \ov \vep^6 & \vep^2 \\
    \end{array}\!\right)
    \!\!\left(\!\!
      \begin{array}{c}
        d^c_1 \\
        d^c_2 \\
        d^c_3 \\
      \end{array}\!\!\right)\!\!\tl \varphi ~+
$$
\beq
   \left(\!\!
  \begin{array}{ccc}
    l_1, &\!\!l_2, &\!\!l_3 \\
  \end{array}\!\!\right)
  \!\!\left(\!
    \begin{array}{ccc}
     \vep^6 & \ov \vep^{38} &\ov \vep^{61}  \\
      \vep^{48} &\vep^4 &\ov \vep^{19} \\
     \vep^{69} & \vep^{25} &\vep^2 \\
    \end{array}\!\right)
    \!\!\left(\!\!
      \begin{array}{c}
        e^c_1 \\
        e^c_2 \\
        e^c_3 \\
      \end{array}\!\!\right)\!\!\tl \varphi ~+{\rm h.c.}
\la{ch-ops}
\eeq
In front of each operator of (\ref{ch-ops}) the dimensionless coupling (omitted here) should stand.
Substituting the VEVs $\lan \vep\ran =\lan \ov \vep \ran \equiv \ep $,
and omitting those terms, with high powers of $\ep$, which are irrelevant in practice,
the
Yukawa matrices $Y_U, Y_D, Y_E$ corresponding to up, down quarks, and charged leptons, respectively, are

\beq
 Y_U\simeq \left(\begin{array}{ccc}

 a_1'\ep^8& ~~ a_1\ep^5& ~~ 0
\\
0&~~ a_2\ep^4 & ~~ \ep^2
 \\
0 & ~~ 0  &~~ 1
\end{array}\right)\lam_t^0~,
\label{UpY}
\eeq
\beq
 Y_D\simeq
 \left(\begin{array}{ccc}

 e^{-i\eta_1}& ~~ 0& ~~ 0
\\
 0 &~~ e^{-i\eta_2} & ~~ 0
 \\
0 & ~~ 0 &~~ 1
\end{array}\right)\!\cdot \!
\left(\begin{array}{ccc}

 0& ~~ b_1\ep^3& ~~ 0
\\
 b_1'\ep^3 &~~ b_2\ep^2 & ~~ b_2'\ep^2
 \\
0 & ~~ 0  &~~ 1
\end{array}\right)\!\!\ka_b\ep^2~.
\label{DownY}
\eeq
\beq
 Y_E\simeq \left(\begin{array}{ccc}

 c_1\ep^4& ~~ 0& ~~ 0
\\
 0 &~~ c_2\ep^2 & ~~ 0
 \\
0 & ~0  &~~ 1
\end{array}\right)\!\!\ka_{\tau }\ep^2~.
\label{YE}
\eeq
We have made field phase redefinitions in such a way that, in this basis, the CKM matrix is the unit matrix [it becomes nontrivial after the diagonalization
of $Y_U$ and $Y_D$ of Eqs. (\ref{UpY}) and (\ref{DownY}), respectively].
Also, we have performed $1-3$ and $2-3$ rotation of $d^c$ states in such a way that $3-1$ and $3-2$ entries of $Y_D$ vanishes
(this transformation of the $d^c_{1,2,3}$ states is unobservable in the SM).
Moreover,  $Y_U$ is real, two phases $\eta_{1,2}$ appear in $Y_D$, while  $Y_E$ is real.
 The phases $\eta_{1,2}$ will not contribute to the quark masses, but will be important for the CKM matrix elements.

Starting with the quark sector,
with  proper (and fully natural) selection of input parameters
 we can get desirable values for fermion masses and CKM mixing angles.
 Since the Yukawa matrices are hierarchical, in a pretty good approximation we can derive the following analytic
 expressions:
 \beq
 \lam_t= \lam_t^0[1+{\cal O}(\ep^4)]~,~~~~
 \fr{\lam_u}{\lam_t}\simeq \fr{a_1'\ep^8}{\sqrt{ 1+(a_1\ep/a_2)^2}}~,~~~~\fr{\lam_c}{\lam_t}\simeq a_2\ep^4\sqrt{ 1+(a_1\ep/a_2)^2}~,
 \la{up-yuk-ratio}
 \eeq
 \beq
 \lam_b= \ka_b\ep^2[1+{\cal O}(\ep^4)]~,~~~~
 \fr{\lam_d}{\lam_b}\simeq \fr{b_1b_1'\ep^4}{\sqrt{ b_2^2+(b_1^2+b_1'^2)\ep^2}}~,~~~~\fr{\lam_s}{\lam_b}\simeq \ep^2\sqrt{ b_2^2+(b_1^2+b_1'^2)\ep^2} ~.
 \la{down-yuk-ratio}
 \eeq
For writing down expression of the CKM matrix elements, it is useful to introduce two angles
\beq
\tan \te_u=\fr{a_1}{a_2}\ep\sqrt{1+\ep^4}~,~~~~\tan 2\te_d=\fr{2b_1b_2\ep \sqrt{1+b_2'^2\ep^4}}{b_2^2-(b_1^2-b_1'^2)\ep^2}~,
\la{12-angs}
\eeq
and notations $\sin \te_{u,d}\equiv s_{u,d}$ and  $\cos \te_{u,d}\equiv c_{u,d}$. With these we have
 $$
 |V_{us}|=\left |c_us_de^{i\eta_1}-s_uc_d\fr{(e^{i\eta_2}+b_2'\ep^4)}{\sqrt{1+\ep^4}\sqrt{1+b_2'^2\ep^4}}\right |+{\cal O}(\ep^7)~,
 $$
 \beq
 |V_{cb}|= c_u\ep^2\fr{\left |1-e^{i\eta_2}b_2'(1+b_2^2\ep^4)\right |}{\sqrt{1+\ep^4}\sqrt{1+b_2'^2\ep^4}}
 +{\cal O}(\ep^8)~,~~~~~~\fr{|V_{ub}|}{|V_{cb}|}=\tan \te_u ~.
 \la{analit-CKM}
 \eeq
For the parameter
\beq
 \ov{\rho}+i\ov{\eta}=-\fr{V_{ud}V_{ub}^*}{V_{cd}V_{cb}^*},
 \la{ro-eta-def}
 \eeq
related to the $CP$ violation and defined in a phase convention independent way \cite{Workman:2022ynf}, we obtain
 \beq
 \ov{\rho}+i\ov{\eta}\simeq \fr{c_uc_de^{i\eta_1}+s_us_de^{i\eta_2}}{c_ds_ue^{i\eta_1}-c_us_de^{i\eta_2}}\tan \te_u  .
 \la{ro-eta-pred}
 \eeq

Upon parametrization of the Yukawa matrices we have taken away the factors $\lam_t^0$ and $\kappa_b$. These will be selected in such
a way as to get observed values of masses $M_t$ and $m_b$. Remaining parameters
 (i.e. $\ep, a_{1,2}, a_1', b_{1,2}, b_{1,2}', \eta_{1,2}$)  will determine light quark masses and CKM matrix elements.
 Relations (\ref{up-yuk-ratio})-(\ref{analit-CKM}) and (\ref{ro-eta-pred}) will help to find parameters giving desirable fit.
Before going to that, let us mention that all quantities (output observable), obtained at high scale $\La $
 (which we take close to the GUT scale - few$\times 10^{16}$~GeV), need to be renormalized at low energies.
For this we  perform the renormalization and calculate these quantities at low scales. We have
$$
\left. \fr{\lam_{u,c}}{\lam_t}\right |_{M_t}=\eta_{u,c}\left. \fr{\lam_{u,c}}{\lam_t}\right |_{\La }~,~~~
\left. \fr{\lam_{d,s}}{\lam_b}\right |_{M_t}=\eta_{d,s}\left. \fr{\lam_{d,s}}{\lam_b}\right |_{\La }~,~~~
$$
$$
\left. V_{\al \bt}\right |_{M_Z}=\eta_{mix}\left. V_{\al \bt}\right |_{\La }~,~~~{\rm if}~~~(\al \bt )=(ub, cb, td, ts)~,
$$
\beq
\left. V_{\al \bt}\right |_{M_Z}=\left. V_{\al \bt}\right |_{\La }~,~~~{\rm if}~~~(\al \bt )=(ud, us, cd, cs, tb)~.
\la{RG-MG-mt}
\eeq
In one-loop approximation we have $\eta_{u,c}\simeq 1/\eta_{d,s}\simeq 1/\eta_{mix}\simeq
\exp \!\l \!\fr{3}{32\pi^2}\!\!\int_{\!M_t}^{\La }\hs{-0.1cm}\lam_t^2d \ln \mu \!\r $. However, we will perform more accurate calculations.
For the renormalization of the light family Yukawa couplings and $\lam_{b, \tau }$
 we use  two-loop renormalization group (RG) equations,  while the runnings of $\lam_t$ and $g_3$ are performed through three-loop RGs.
 For the running of the CKM matrix  elements the two-loop RGs \cite{Barger:1992pk} will be used.
 Upon the running between $M_t$ (the pole mass of the top quark) and the scale $\La $, for boundary values
 of the couplings at $\mu =M_t$ we use values given in \cite{Martin:2022qiv}.

Doing so, for  $M_t=172.5$~GeV and $\al_3(M_Z)=0.1179$ (the values we use throughout of this work) we get
\beq
\eta_{u,c}\simeq 1.1262+0.00187\cdot \ln \!\l\!\fr{\La}{2\!\cdot \!10^{16}{\rm GeV}}\!\r ,
\la{RG-uc}
\eeq
\beq
\eta_{d,s}\simeq 0.8916-0.00143\cdot \ln \!\l\!\fr{\La}{2\!\cdot \!10^{16}{\rm GeV}}\!\r ,
\la{RG-ds}
\eeq
\beq
\eta_{mix}\simeq 0.89157- 0.001433\cdot \ln \!\l\!\fr{\La}{2\!\cdot \!10^{16}{\rm GeV}}\!\r ,
\la{RG-CKM}
\eeq
which are the interpolated expressions that work pretty well for $10^{15} {\rm GeV}\!<\!\La \!<\!M_{\rm Pl}$.

Also, for light quark masses, the running from $M_t$ down to low scales need to be performed by the standard technics
 \cite{Martin:2022qiv,Fusaoka:1998vc,Xing:2007fb}.

\subsubsection*{A. Fit for charged Fermion masses and CKM elements}

A good fit is obtained for the following values of input parameters (values are given at high scale $\La =2\tm 10^{16}$~GeV):
$$
\ep =0.21,~~\{a_1, a_1', a_2\}=\{0.6974,~ 1.7065, ~1.6606 \},~~\{\eta_1, \eta_2\}=\{3.01985,~ -1.3954 \},
$$
\beq
\{b_1, b_1', b_2, b_2'\}=\{0.47834,~ 0.54541,~ 0.45448,~ 0.59088\}.
\la{inp}
\eeq
These, by performing renormalization [using (\ref{RG-MG-mt})-(\ref{RG-CKM}) and input $M_t=172.5$~GeV, $m_b(m_b)=4.18$~GeV],
at low scales give

$$
 \l m_u,~ m_d,~ m_s \r(2~{\rm GeV})=\l 2.16,~ 4.67, ~93\r {\rm MeV},~~~m_c(m_c)=1.27 ~{\rm GeV}~,
$$
 \beq
 {\rm at}~~\mu=M_Z:~~|V_{us}|=0.225,~~|V_{cb}|=0.04182,~~|V_{ub}|=0.00369,~~~\ov{\rho}=0.159,~~\ov{\eta}=0.3477,
 \la{low-sc}
 \eeq
 where definitions for  $\ov{\rho}, \ov{\eta}$ are given in Eq. (\ref{ro-eta-def}).
All results given above are in perfect agreement with experiments \cite{Workman:2022ynf}.

As far as the charged lepton masses are concerned, from (\ref{YE})
with the input $M_{\tau }=1.777$~GeV and
\beq
{\rm at}~~\mu=\La ~,~~~~\{c_1, c_2\}\simeq \{0.1437, 1.335\},
\la{E-GUT}
\eeq
and taking into account that
$\left. \fr{\lam_{e,\mu}}{\lam_{\tau}}\right |_{M_Z}\!\!\cong \!\left. \fr{\lam_{e,\mu}}{\lam_{\tau}}\right |_{\La }$,
 we obtain
\beq
M_e= 0.511~{\rm MeV},~~~~M_{\mu }=105.66~{\rm MeV},
\la{lept-mass-low}
\eeq
which is also in agreement with experiments.

\section{NEUTRINO SECTOR}
\la{nu-sect}

For building the realistic neutrino sector, the singlet states $N_{1,2,3}$ will be used as right-handed neutrinos.
Since the $N_4$ is not really needed for these purposes, its couplings to the leptons and also to $N_{1,2,3}$
can be easily avoided by imposing the reflection symmetry $N_4\to -N_4$ (this symmetry and
its possible implication will be commented on below). This will make the analysis simpler.
Thus, with $U(1)_F$ charges given in Table \ref{t:tab1} the $lN\varphi $ and $N_iN_j$ type couplings ($i,j=1,2,3$)
 will be
\beq
\left(\!\!
  \begin{array}{ccc}
    l_1, &\!\!l_2, &\!\!l_3 \\
  \end{array}\!\!\right)
  \!\!\left(\!
    \begin{array}{ccc}
      \ov \vep^9 & \ov \vep^{51} & \ov \vep^{52} \\
     \vep^{33} &\ov \vep^9 &\ov \vep^{10} \\
      \vep^{54} &  \vep^{12} & \vep^{11} \\
    \end{array}\!\right)
    \!\!\left(\!\!
      \begin{array}{c}
        N_1 \\
        N_2 \\
        N_3 \\
      \end{array}\!\!\right)\!\!\varphi +
   \left(\!\!
  \begin{array}{ccc}
    N_1, &\!\!N_2, &\!\!N_3 \\
  \end{array}\!\!\right)
  \!\!\left(\!
    \begin{array}{ccc}
       \vep^{64} & \vep^{22} & \vep^{21} \\
      \vep^{22} & \ov \vep^{20} &\ov \vep^{21} \\
      \vep^{21} & \ov \vep^{21} & \ov \vep^{22} \\
    \end{array}\!\right)
    \!\!\left(\!\!
      \begin{array}{c}
        N_1 \\
        N_2 \\
        N_3 \\
      \end{array}\!\!\right)\!\! M_{\rm Pl}.
\label{Nu-coupl}
\eeq
In these operators the dimensionless couplings are still omitted.
Substituting the VEVs $\lan \vep \ran =\lan \ov \vep \ran =\ep $, $\lan \varphi \ran =v$ and omitting irrelevant small entries, for neutrino Dirac
and Majorana matrices we get
\beq
m_D\!\simeq \!\left(
      \begin{array}{ccc}
        A\ep^9 & 0 & 0 \\
        0 & B_1\ep^9 & C_1\ep^{10} \\
        0 & B_2\ep^{12} & C_2\ep^{11} \\
      \end{array}
    \right)\!\!v ,~~~
    M_R\!\simeq \!  \left(
                      \begin{array}{ccc}
                        0 & a\ep^2 & d\ep \\
                        a\ep^2 & b & c\ep \\
                        d\ep & c\ep & \ep^2 \\
                      \end{array}
                    \right)\!\!\bar c M_{Pl}\ep^{20}.
\la{mD-MR}
\eeq
These lead to the light neutrino $3\tm 3$ mass matrix:
\beq
\begin{array}{ccc}
 & {\begin{array}{ccc}
\hs{-0.6cm} &~~  &~~ \hs{0.2cm}
\end{array}}\\ \vspace{1mm}
\begin{array}{c}
 \\  ~  \\  ~
 \end{array}\!\!\!\!\!\hs{-0.1cm}&{\!  M_{\nu }\simeq -m_DM_R^{-1}m_D^T\simeq \left(\begin{array}{ccc}

\bt & ~~\ga  & ~~\ga'
\\
 \ga &~~ \al^2 & ~~\al
 \\
\ga'& ~~\al &~~ 1
\end{array}\right)\bar m},
\end{array}  \!\!  ~~~
\label{nu-matrix1}
\eeq
with $\bar m$ and the dimensionless couplings $\al ,\bt ,\ga ,\ga'$ expressed by the scales and couplings appearing in
Eq. (\ref{mD-MR}). Note that
$M_{\nu }$'s $2-3$ block's determinant is zero:
\beq
 M_{\nu }^{(2,2)} M_{\nu }^{(3,3)}-(M_{\nu }^{(2,3)})^2= 0.
\la{det23-0}
\eeq
The origin of this relation can be understood as follows. Because of $M_R^{(1,1)}=0$, the determinant of the lower $2\tm 2$ block of $M_R^{-1}$ is zero.
Moreover, since the lower $2\tm 2$ block of $m_D$ decouples [i.e.,  $(1,2)$ and $(1,3)$ entries in $m_D$ are zero] the seesaw
formula $M_{\nu }\simeq -m_DM_R^{-1}m_D^T$ gives the relation of Eq. (\ref{det23-0}).
The latter gives specific predictions, on which we will focus now.

Since the charged lepton mass matrix $Y_E$ is essentially diagonal, the whole lepton mixing matrix $U$ comes from the neutrino sector.
Therefore, we have
\beq
 M_{\nu }=PU^*P'M_{\nu}^{\rm Diag}U^{\dag }P,
\la{barMnu}
\eeq
where  in a standard parametrization, $U$ has the following form:
\begin{equation}
U=
\left(\begin{array}{ccc}c_{13}c_{12}&c_{13}s_{12}&s_{13}e^{-i\delta}\\
-c_{23}s_{12}-s_{23}s_{13}c_{12}e^{i\delta}&c_{23}c_{12}-s_{23}s_{13}s_{12}e^{i\delta}&s_{23}c_{13}\\
s_{23}s_{12}-c_{23}s_{13}c_{12}e^{i\delta}&-s_{23}c_{12}-c_{23}s_{13}s_{12}e^{i\delta}&c_{23}c_{13}
\end{array}\right),
\la{Ulept}
\end{equation}
with $s_{ij}=\sin\theta_{ij}$ and $c_{ij}=\cos\theta_{ij}$. The phase matrices $P, P'$ are given by
\beq
P={\rm Diag}\l e^{i\om_1}~,~e^{i\om_2}~,~e^{i\om_3}\r,~~~
P'={\rm Diag}\l 1~,~e^{i\rho_1}~,~e^{i\rho_2}\r,
\la{Ps}
\eeq
where $\om_{1,2,3}, \rho_{1,2}$ are some phases.

As was investigated in details (see second Ref. in \cite{Tavartkiladze:2011ex}), the relation (\ref{det23-0}) excludes the possibility of the normal ordering of the neutrino
masses.
Using Eqs. (\ref{barMnu})-(\ref{Ps}) in (\ref{det23-0}) the   we obtain
\beq
e^{i\rho_1}m_1m_2s_{13}^2+e^{i(\rho_2+2\de)}\l m_1 s_{12}^2+e^{i\rho_1}m_2 c_{12}^2\r m_3c_{13}^2=0,
\la{pred-rel}
\eeq
which in turn gives
\beq
\cos \rho_1=\fr{m_1^2m_2^2\tan^4\te_{13}-m_3^2(m_1^2s_{12}^4+m_2^2c_{12}^4)}{2m_1m_2m_3^2s_{12}^2c_{12}^2},
\la{nu-pred1}
\eeq
\beq
2\de =\pm \pi-\rho_2+{\rm Arg}\l s_{12}^2e^{i\rho_1}+\fr{m_2}{m_1}c_{12}^2\r.
\la{nu-pred2}
\eeq
These two relations, together with measured values of $\De m_{\rm sol}^2$ and $\De m_{\rm atm}^2$ allow us to have only one free phase
(out of the three phases $\de , \rho_{1,2}$) and one free mass (out of the three light neutrino masses $m_{1,2,3}$).
However, as we will see below, the  latter's range will turn out to be quite narrow.

Using recent results from the neutrino experiments
\cite{Capozzi:2021fjo}, we can easily verify
that the relation of Eq. (\ref{nu-pred1}) is incompatible with normal ordering of neutrino masses.
On the other hand, inverted ordering  of neutrino masses is possible.
Using the best fit values (bfvs) of $\te_{ij}, \De m_{\rm sol}^2=m_2^2-m_1^2$, $\De m_{\rm atm}^2=m_2^2-m_3^2$,
 expressing $m_{1,2}$ by $m_3$ as $m_1\!=\!\sqrt{\De m_{\rm atm}^2\!-\!\De m_{\rm sol}^2\!+\!m_3^2}$, $m_2\!=\!\sqrt{\De m_{\rm atm}^2\!+\!m_3^2}$,
 from Eq.  (\ref{nu-pred1})
we can get an allowed region for $m_3$:
\beq
0.001129~{\rm eV}\stackrel{<}{_\sim }m_3\stackrel{<}{_\sim }0.002833~{\rm eV}.
\la{predictions}
\eeq
This implies $0.1002~{\rm eV}\stackrel{<}{_\sim }\sum m_i\stackrel{<}{_\sim } 0.1021$~eV, satisfying the current upper bound
$\sum m_i<0.12$~eV \cite{Planck:2018vyg},  which is obtained from
cosmology.

\begin{figure}[!t]
\begin{center}
\includegraphics[width=0.8\columnwidth]{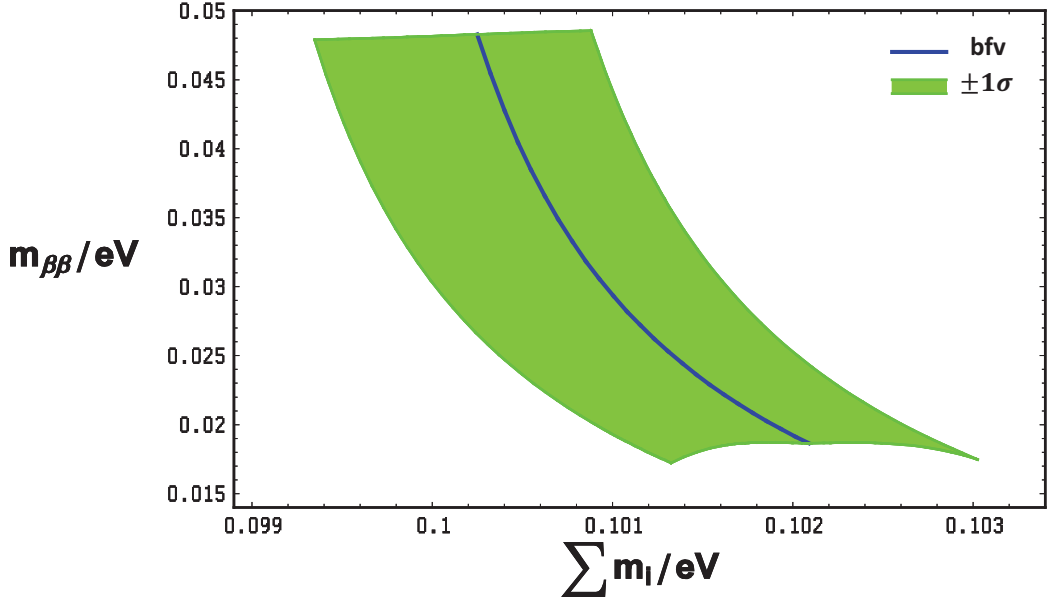}
\vs{-0.3cm}
\caption{Correlation between  $\sum m_i$ and $m_{\bt \bt}$. Solid (middle) blue line corresponds to the bfvs of the oscillation
parameters \cite{Capozzi:2021fjo}.
Green (wider) area corresponds to the cases with oscillation parameters within the $1\si $ deviations.
}
\vs{-0.3cm}
\label{fig1}
\end{center}
\end{figure}

 Moreover, for neutrino less double $\bt $-decay ($0\nu \bt \bt $)
parameter $m_{\bt \bt}=|\sum U_{ei}^2m_iP^{\!\!~'\!*}_i|$ we obtain
\beq
m_{\bt \bt}= \left |c_{12}^2c_{13}^2m_1+s_{12}^2c_{13}^2m_2e^{i\rho_1}+s_{13}^2m_3e^{i(2\de+\rho_2)} \right |,
\la{2bt-decay}
\eeq
which taking into account Eqs. (\ref{nu-pred1}), (\ref{nu-pred2}) and bfvs of the oscillation parameters leads to:
\beq
0.01864~{\rm eV}\stackrel{<}{_\sim }m_{\bt \bt}\stackrel{<}{_\sim }0.0483~{\rm eV}.
\la{predictions1}
\eeq
This range is also compatible with limits provided by $0\nu \bt \bt $ experiments \cite{KamLAND-Zen:2022tow}.
In fact, due to predictive relations  in Eqs. (\ref{nu-pred1}), (\ref{nu-pred2}) both parameters $\sum m_i$ and $m_{\bt \bt}$ are
unequivocally  determined by the $m_3$ values.\footnote{Note that, thanks to the relations of (\ref{nu-pred1}) and (\ref{nu-pred2}),
 the phase $\rho_1$ and the combination $2\de+\rho_2$ entering in (\ref{2bt-decay}) are
 unequivocally  determined by the $m_3$.} Thus, there is correlation between
 $\sum m_i$ and $m_{\bt \bt}$, which is given in
Fig. \ref{fig1}.
Hopefully, future experiments will be able to test viability of this scenario \cite{Abazajian:2022ofy}.
%
%

Now we give one selection of the parameters, appearing in (\ref{mD-MR}), which blends well with this neutrino scenario
and then discuss some implications and outcomes.
With the choice
$$
\{a, b, c, d, \bar c\}\simeq \{3.2672e^{i1.5473}, 0.79405e^{i0.0053733}, 0.89097e^{i0.0028735}, 0.15853e^{1.5586}, 0.56333e^{2.9194}  \},
$$
\beq
\{ A, B_1, B_2, C_1, C_2\}\simeq \{2.0236,~ 2.0236,~ 1.6189,~ 2.4283,~ -0.8094\},
\la{inp-pars}
\eeq
for the light neutrino masses and mixing angles we obtain
\beq
\{m_1, m_2, m_3 \}=\{0.049197, ~0.049942,~ 0.0015\}{\rm eV},
\la{nu-masses}
\eeq
\beq
\{\sin^2\te_{12},  \sin^2\te_{23}, \sin^2\te_{13}\}=\{0.3035,~ 0.57,~ 0.02235\}.
\la{nu-mixings}
\eeq
{}From (\ref{nu-masses}) we get
\beq
\De m_{\rm sol}^2=m_2^2-m_1^2=7.39\tm 10^{-5}{\rm eV}^2,~~
\De m_{\rm atm}^2=m_2^2-m_3^2= 2.492\tm 10^{-3}{\rm eV}^2~.
\la{nu-mass2dif}
\eeq
Results of (\ref{nu-mixings}) and (\ref{nu-mass2dif})
correspond to the bfvs of the inverted ordering neutrino scenario \cite{Capozzi:2021fjo}.
Moreover, for the phases we get
\beq
\{\de, \rho_1, \rho_2 \}=\{276^\circ ,~ 91.69^\circ ,~ 11.49^\circ \},~~~~\om_{1,2,3}=0.
\la{phases}
\eeq
For this case we have $m_{\bt \bt}\simeq 0.0362$~eV and $\sum m_i\simeq 0.101$~eV. These certainly blend with the discussed
predictions [of Eqs. (\ref{nu-pred1}), (\ref{nu-pred2}) and Fig. \ref{fig1}].

From the input (\ref{inp-pars}) for the heavy RHN masses we get
\beq
\{M_{N_1}, M_{N_2}, M_{N_3} \}\simeq \{1.6,~953.5, ~32480 \}{\rm GeV}.
\la{N-masses}
\eeq
A few remarks about the heavy RHN sector are in order.

The state $N_1$ (with $M_{N_1}\simeq 1.6$~GeV) can be produced in decays of heavy mesons; however
the corresponding mixing $|U_{eN_1}|^2\simeq 2.76\tm 10^{-11}$ is a bit below (by factor of$\approx 3$) the sensitivity
of the SHiP experiment \cite{SHiP:2018xqw}. As a separate study, it
would be interesting to do a more detailed investigation/exploration of the model's parameters from the perspective of this
experiment.

Since the lightest RHN's mass is $M_{N_1}\simeq 1.6$~GeV, and it
 mixes with $\nu_e$, there will be an additional  contribution to the $02\bt \bt $ parameter, which is given by  \cite{Mitra:2011qr}
$$
\left |\sum_{i=1}^3 U_{ei}^2m_iP^{\!\!~'\!*}_i +\fr{M_{N_1}}{1+M_{N_1}^2/\lan p^2\ran }U_{eN_1}^2 \right |=
$$
\beq
\left | e^{-0.421i}0.0362~\! {\rm eV}+\fr{e^{-0.151i}2.76\cdot 10^{-11}M_{N_1}}{1+M_{N_1}^2/\lan p^2\ran } \right |\!=\!0.0368~ {\rm eV},
~\l {\rm for}~ \lan p^2\ran\!=\!(200~\!{\rm MeV})^2\r .
\la{2bt-N1}
\eeq
The second term in the absolute values of Eq. (\ref{2bt-N1}) is the contribution from the $N_1$. The $\lan p^2\ran $ is
averaged momentum squared corresponding to this process.
As can be seen, for $\lan p^2\ran \!=\!(\!100\!-\!200~\!{\rm MeV})^2$ \cite{Mitra:2011qr, Bolton:2022pyf}
the correction from the $N_1$ state is within $(0.5\!-\!1.8)\%$, i.e. negligible. Therefore, the predictions, made from the
light neutrino sector (and correlation of Fig. \ref{fig1}) are robust.

With $N_1$'s mass within the GeV scale,
we need to ensure its sufficiently fast decay (within $\stackrel{<}{_\sim }0.3$ sec.)  in order to not affect the standard big bang
nucleosynthesis (BBN).
Dominant decays of $N_1$ are three body decays via neutral and charged currents (i.e., via $Z^*$ and $W^{*(\pm )}$ exchange).
These are leptonic $N_1\to \nu_i \nu_j\bar \nu_j, ~\nu_i e^{-}_je^{+}_j,~ e^{-}_ie^{+}_j\nu_j$ and semileptonic
$N_1\to \nu_i q_j\bar q_j,~ e^{-}_iu_j\bar d_k$ decays.
For the leptonic decay widths we use expressions given in Ref. \cite{Atre:2009rg}.
For the semileptonic decays, taking into account all inclusive decays into the quarks, by proper use of the matching RG factor
\cite{Bondarenko:2018ptm} one can get a quite reasonable estimate.
Summing all kinematically allowed channels of $N_1$'s decays and using proper expressions \cite{Atre:2009rg, Bondarenko:2018ptm},
for the total width (i.e., for inverse lifetime) we obtain
 \beq
\Ga (N_1)=\fr{1}{\tau_{N_1}}
\!\simeq \!\fr{G_F^2M_{N_1}^5}{16\pi^3}\!\l \!
1.37|U_{1N_1}|^2\!+\!1.35|U_{2N_1}|^2\!+\!0.487|U_{3N_1}|^2
~\!\r \simeq \fr{1}{0.0038~\!{\rm s.}}~,
\la{N1-decay}
\eeq
which is compatible with BBN.
In Eq. (\ref{N1-decay}), for the squared mixing matrix elements we have used values obtained within our model:
\beq
|U_{iN_1}|^2\simeq \{2.76,~ 1.29,~1.09 \}\tm 10^{-11},
\la{UiN-values}
\eeq
which obtained from the inputs of (\ref{inp-pars}).
The states $N_{2,3}$ will decay much rapidly via the $N_{2,3}\to \varphi l$ channel (with lifetimes $\approx 7\tm 10^{-3}$~ps and
$2\tm 10^{-4}$~ps respectively).
As far as the state $N_4$  (which presence is important for anomaly cancelation) is concerned, because of the reflection symmetry  $N_4\to -N_4$ (we have
introduced), its mixing with $N_{1,2,3}$ and couplings to the SM leptons are forbidden.
However, it will gain the mass via the  $\fr{1}{2}M_{Pl}\bar \vep^{10}N_4N_4$  operator: $M_{N_4}\sim M_{Pl}\ep^{10}\approx 4\tm 10^{11}$~GeV.
For its decay are responsible the operators
\beq
\fr{\lam_1 \bar \vep}{M_{\rm Pl}^2}(N_4u^c_3)(d^c_1d^c_2)+
\fr{\lam_2\vep}{M_{\rm Pl}^2}(N_4u^c_2)(d^c_1d^c_3) +{\rm h.c.},
\la{N4-ops}
\eeq
which are allowed if all quarks also change sign. [I.e., $(q, u^c, d^c)\to -(q, u^c, d^c)$ under reflection symmetry. This does not affect the charged fermion and neutrino sectors.]
These operators will give decays $N_4\to u^c_3d^c_1d^c_2,~  u^c_2d^c_2d^c_3$.
Since $N_4$ is a Majorana state, also $N_4\to \bar u^c_3\bar d^c_1\bar d^c_2,~  \bar u^c_2\bar d^c_2\bar d^c_3$ decays will proceed.
All these give
 $\Ga(N_4)\!=\!\fr{(|\lam_1|^2+|\lam_2|^2)M_{N_4}^5\ep^2}{128\pi^3M_{\rm Pl}^4}
 \!= \!\fr{1}{10^{-4}{\rm sec.}}\l \!\fr{M_{N_4}}{4\cdot 10^{11}{\rm GeV}}\!\r^{\!5}$ (with $\lam_{1,2}=1$),
 and therefore making  $N_4$ harmless for the BBN. It would have been interesting to have a scenario with $N_4$ having proper value of
 mass and needed couplings for serving as a dark matter candidate. This turned out impossible with a presented $U(1)_F$ charge assignment.
 Perhaps a separate study  focused on this issue is also worthwhile.

In summary, exploring the possibility of anomaly free gauged $U(1)_F$ flavor symmetry offered an
attractive pattern for the charged fermion masses, neutrino oscillations, and also interesting
phenomenological implications. These motivate us to think more and try to find other possibilities
within the framework discussed in this work.

\bibliographystyle{unsrt}

\begin{thebibliography}{99}


\bibitem{Capozzi:2021fjo}
F.~Capozzi, E.~Di Valentino, E.~Lisi, A.~Marrone, A.~Melchiorri and A.~Palazzo,
Phys. Rev. D \textbf{104},  083031 (2021);
%
M.~C.~Gonzalez-Garcia, M.~Maltoni and T.~Schwetz,
Universe \textbf{7},  459 (2021).


\bibitem{Froggatt:1978nt}
  C.~D.~Froggatt and H.~B.~Nielsen,
   Nucl.\ Phys.\ B{\bf 147}, 277 (1979).






\bibitem{Dudas:1995yu}
  E.~Dudas, S.~Pokorski, and C.~A.~Savoy,
   Phys.\ Lett.\ B {\bf 356}, 45 (1995).

\bibitem{Chen:2008tc}
  M.~-C.~Chen, D.~R.~T.~Jones, A.~Rajaraman, and H.~-B.~Yu,
   Phys.\ Rev.\ D {\bf 78}, 015019 (2008).



\bibitem{Tavartkiladze:2011ex}
  Z.~Tavartkiladze,
  Phys.\ Lett.\  B {\bf 706}, 398 (2012);
%
%
Phys. Rev. D \textbf{87}, 075026 (2013).


\bibitem{Barger:1992pk}
V.~D.~Barger, M.~S.~Berger, and P.~Ohmann,
Phys. Rev. D \textbf{47}, 2038 (1993).

\bibitem{Martin:2022qiv}
S.~P.~Martin,
Phys. Rev. D \textbf{106},  013007 (2022).
See also references therein and Ref. \cite{Xing:2007fb} for the review and summary of RG equations.

\bibitem{Fusaoka:1998vc}
  H.~Fusaoka and Y.~Koide,
  Phys.\ Rev.\  D {\bf 57}, 3986 (1998).


\bibitem{Xing:2007fb}
Z.~z.~Xing, H.~Zhang and S.~Zhou,
Phys. Rev. D \textbf{77}, 113016 (2008);
%
%
Z.~z.~Xing,
Phys. Rep. \textbf{854}, 1 (2020).


\bibitem{Workman:2022ynf}
R.~L.~Workman (Particle Data Group),
Prog. Theor. Exp. Phys. \textbf{2022}, 083C01 (2022).


\bibitem{Planck:2018vyg}
N.~Aghanim \textit{et al.} (Planck Collaboration),
Astron. Astrophys. \textbf{641}, A6 (2020);
\textbf{652}, C4(E) (2021).


\bibitem{KamLAND-Zen:2022tow}
S.~Abe \textit{et al.} (KamLAND-Zen Collaboration),
arXiv:2203.02139.




\bibitem{Abazajian:2022ofy}
K.~N.~Abazajian, N.~Blinov, T.~Brinckmann, M.~C.~Chen, Z.~Djurcic, P.~Du, M.~Escudero, M.~Gerbino, E.~Grohs, S.~Hagstotz \textit{et al.},
arXiv:2203.07377.




\bibitem{SHiP:2018xqw}
C.~Ahdida \textit{et al.} (SHiP Collaboration),
J. High Energy Phys. \textbf{04} (2019) 077.



\bibitem{Mitra:2011qr}
M.~Mitra, G.~Senjanovic, and F.~Vissani,
Nucl. Phys. B\textbf{856}, 26 (2012).

\bibitem{Bolton:2022pyf}
%
P.~D.~Bolton, F.~F.~Deppisch, and P.~S.~Bhupal Dev,
J. High Energy Phys. \textbf{03} (2020) 170.
%
%



\bibitem{Atre:2009rg}
A.~Atre, T.~Han, S.~Pascoli, and B.~Zhang,
J. High Energy Phys. \textbf{05} (2009) 030.


\bibitem{Bondarenko:2018ptm}
K.~Bondarenko, A.~Boyarsky, D.~Gorbunov, and O.~Ruchayskiy,
J. High Energy Phys. \textbf{11}, (2018) 032, see also references therein.
\end{thebibliography}

\end{document}